\documentclass{osa-article}

\journal{osajournal}

\articletype{Research Article}

\let\oldhat\hat
\renewcommand{\vec}[1]{\mathbf{#1}}
\renewcommand{\hat}[1]{\oldhat{\mathbf{#1}}}

\renewcommand{\Re}{\operatorname{Re}}

\begin{document}

\noindent\fbox{\texttt{version: \today}}\vspace{1cm}

\title{Frequency-domain calculation of Smith-Purcell radiation for metallic and dielectric gratings}

\author{
Andrzej Szczepkowicz,\authormark{1} 
Levi Sch\"{a}chter,\authormark{2}
Joel England\authormark{3}
}

\address{
\authormark{1}Institute of Experimental Physics, University of Wroclaw, Plac M.~Borna 9, 50-204 Wroclaw, Poland\\
\authormark{2}Technion--Israel Institute of Technology, Haifa 32000, Israel\\
\authormark{3}SLAC National Accelerator Laboratory, 2575 Sand Hill Road, Menlo Park, CA 94025, USA
}

\begin{abstract}
The intensity of Smith-Purcell radiation from metallic and dielectric gratings (silicon, silica) is compared in a frequency-domain simulation. The numerical model is discussed and verified with the Frank-Tamm formula for Cherenkov radiation. For 30 keV electrons, rectangular dielectric gratings are less efficient than their metallic counterpart, by an order of magnitude for silicon, and two orders of magnitude for silica. For all gratings studied, radiation intensity oscillates with grating tooth height due to electromagnetic resonances in the grating. 3D and 2D numerical models are compared.
\end{abstract}

\section{Introduction}

The Smith-Purcell (SP) radiation, observed for visible light in 1953 \cite{1953-Smith-Purcell}, has been shown to occur in a wide spectral region, from microwaves \cite{1969-Rusin-Bogomolov,2004-Grishin-Fuchs} generated using macroscopic gratings, to ultraviolet radiation \cite{2014-So-MacDonald,2019-Ye-Liu} from nanogratings.
One foreseen application of this effect would be a highly tunable free-electron light source \cite{2019-RoquesCarmes-Kooi}.
SP radiation might also be used for beam diagnostics in accelerators,
for beam position monitoring \cite{1989-Fernow, 2001-Doucas-Kimmitt,2012-Soong-Byer} or longitudinal profile characterization \cite{1997-Lampel, 1997-Nguyen, 2002-Doucas-Kimmitt, 2009-Blackmore-Doucas, 2012-Bartolini-Clarke}.
New motivation to study SP radiation comes from the development of Dielectric Laser Accelerators (DLA) \cite{2014-England-Noble,2020-Sapra-Yang}, which utilize the
inverse Smith-Purcell effect. Electron beams from DLA may in turn be used to generate SP radiation in various spectral regions.

The majority of experimental studies of SP radiation were carried out with metallic gratings. Some recent studies deal with dielectric gratings (eg.~\cite{2018-Yang-Massuda, 2019-RoquesCarmes-Kooi}); this is caused by advances in dielectric nanofabrication, improved understanding of SP emission from dielectrics, and hope that dielectrics may in some cases outperform metals in radiation intensity \cite{2019-RoquesCarmes-Kooi}.

Calculations of SP radiation intensity from gratings have a long history. Most analytical work to date considers only metallic gratings
(exception: a very simplified model in Ref.~\cite{2010-Sukhikh-Naumenko} applied to sub-THz radiation). 
Ref.~\cite{2011-Potylitsyn} reviews some of the analytical models,
and their outcomes are compared in Refs~\cite{2006-Karlovets-Potylitsyn, 2016-Malovytsia-Delerue}. Many of the models build upon the seminal work by
Toraldo di Francia \cite{1960-diFrancia}, which treats both SP and Cherenkov radiation with the same formalism (``Cherenkovian effects''). The range of analytical methods include perturbative approaches valid for shallow gratings \cite{1966-Barnes-Dedrick,1973-Lalor} and
various surface current models \cite{1998-Brownell-Walsh, 2000-Trotz-Brownell, 2002-Kube-Backe, 2005-Brownell-Doucas, 2011-Potylitsyn} which are best suited for shallow gratings, but for high energies can also be applied to deep profiles \cite{1998-Brownell-Walsh}. 
Most of the analytical models involve some approximations and neglect resonant cavity effects in the grating. 
According to Ref.~\cite{2006-Karlovets-Potylitsyn} the results of different analytical models may differ by up to six orders of magnitude.
An exceptional position among the analytical models of SP radiation is held by the van den Berg’s model 
\cite{1973-vandenBerg2D, 1973-vandenBerg3D, 1974-vandenBerg-Tan}. According to the author the model is rigorous and is applicable to arbitrary grating profile. Although the model's accuracy has been questioned \cite{1998-Brownell-Walsh}, no one has explicitly shown the model to be inexact. The model does reproduce radiation intensity oscillations with increasing tooth height \cite{1998-Shibata-Hasebe}, a resonant cavity effect in the grating. However, van den Berg's approach is probably the most difficult of the SP models to apply and in the end requires nontrivial numerical calculations \cite{1973-vandenBerg2D,1994-Haeberle-Rullhusen,1998-Brownell-Walsh}.

Metals are easier to deal with in analytical models than dielectrics, because with the perfect electric conductor boundary condition it is not necessary to solve for the field inside and on the other side of the grating.
Regarding dielectric gratings, Sukhikh et al.\ \cite{2010-Sukhikh-Naumenko} report analytical calculation of SP radiation from a teflon grating, however with several special assumptions: geometry of an inverted lamelar grating (rectangular grating) with substrate thickness approaching zero, SP radiation only to one side, and neglect of secondary refractions (resonances within the grating are not reproduced); the model has been applied in \cite{2010-Sukhikh-Naumenko} for sub-THz radiation.

In recent years an increasing number of purely numerical simulations of SP radiation
were reported. Numerical simulations are equally applicable to metallic and dielectric gratings, although for a dielectric grating more time and memory resources are needed. The most common approach is a time-domain simulation; some recently used solvers are Lumerical FDTD \cite{2017-Lai-Kuang,2017-Kaminer-Kooi,2017-Remez-Shapira,2018-Massuda-Roques-Carmes} and CST \cite{2014-Liu-Xu,2015-Lekomtsev-Karataev,2017-Aryshev-Potylitsyn,2019-Zhang-Konoplev}. The time of calculation is usually from hours to days on a single CPU machine. Another approach is a frequency-domain simulation, which is much faster if infinite gratings are assumed (simulation for one unit cell with periodic boundary conditions). This approach was used in Refs.\ 
\cite{2018-Yang-Massuda, 2018-Massuda-Roques-Carmes, 2018-Song-Du}; however, none of these papers describes the method of simulation, and it is hard to deduce how radiation energy was calculated. A description of the simulation method can be found in papers which report frequency-domain calculations of Cherenkov radiation \cite{2018-Tyukhtin-Vorobev, 2018-Galyamin-Tyukhtin, 2019-Galyamin-Vorobev}, but these papers do not compute radiated energy.
All of the Refs. \cite{2018-Yang-Massuda, 2018-Massuda-Roques-Carmes, 2018-Song-Du,2018-Tyukhtin-Vorobev, 2018-Galyamin-Tyukhtin, 2019-Galyamin-Vorobev} use the Comsol frequency-domain solver \cite{Comsol}. 

The present work focuses on frequency-domain simulation of single-electron (``incoherent'') Smith-Purcell radiation with metallic and dielectric gratings. We start in Sect.~\ref{sect-model} with a step-by-step description of calculation method for Smith-Purcell and Cherenkov radiation (``Cherenkovian effects’’ \cite{1960-diFrancia}) using a frequency-domain numerical solver. Although simple in principle, the solution requires careful differentiation between phasors (as required by the numerical solver) and phasor densities (Fourier transforms) and proper interpretation of the well known expression $\Re[\frac{1}{2}\vec E\times \vec H^*]$, which is different for phasors and for phasor densities. Careful treatment leads to the solution that is correct in absolute terms, without spurious multiplicative constants. We verify our method by comparing the results for Cherenkov radiation with the exact analytical Frank--Tamm formula for radiated energy \cite{1937-Frank-Tamm,1998-Jackson,2004-Shiozawa}.

After the detailed deliberations on methodology we turn to applications. In Sect.~\ref{sect-comparison-materials} we use our frequency-domain model to compare directly radiation from gratings of fixed geometry and different materials, which to our knowledge has not yet been reported in the literature, except for the mentioned previously very limited model in Ref.~\cite{2010-Sukhikh-Naumenko}, and except for a recent paper \cite{2019-RoquesCarmes-Kooi}, which however compares theoretical upper bounds for SP radiation, not the actual computed values.
In Sect.~\ref{sect-oscillations} we demonstrate that the model captures resonant effects in the grating. While energy oscillations with increasing tooth height have been reported previously for metallic gratings \cite{1998-Shibata-Hasebe, 2014-Liu-Xu, 2015-Liu-Li,2015-Lekomtsev-Karataev}, we demonstrate them for the first time for dielectrics.
Section \ref{sect-comparison-3d-2d} briefly compares a numerical result from a three-dimensional (3D) and a two-dimensional (2D) model. This is an important issue, as the 3D models require large RAM memory and are more difficult to construct, so one usually starts with 2D modelling.
Section~\ref{sect-triangular} briefly describes radiation from triangular gratings, and Sect.~\ref{sect-summary} summarizes the paper.

In all equations in this paper we use SI units.

\section{\label{sect-model}Calculation of Smith-Purcell or Cherenkov radiation intensity with a frequency-domain solver}

\subsection{\label{sect-phasors-vs-densities}Phasors vs.\ phasor densities and the expression for energy}

To perform calculations using a numerical frequency-domain solver, we must carefully distinguish between phasors and phasor densities.
In case of time-harmonic electromagnetic field we have
\begin{subequations}
	\label{eq-physical-quantities}
\begin{align}
	\vec J(\vec r,t) & 
	=\Re[\vec J(\vec r) e^{j\omega_0 t}] 
	=\frac{1}{2}[\vec J(\vec r) e^{j\omega_0 t}+\vec J^*(\vec r) e^{-j\omega_0 t}]
    \label{eq-pq-J}\\
    \vec E(\vec r,t) & 
	=\Re[\vec E(\vec r) e^{j\omega_0 t}] 
	=\frac{1}{2}[\vec E(\vec r) e^{j\omega_0 t}+\vec E^*(\vec r) e^{-j\omega_0 t}]
	\label{eq-pq-E}
\end{align}
\end{subequations}
and similarly for the $\vec B$, $\vec D$ and $\vec H$ fields. 
The phasors, denoted here $\vec J(\vec r)$, $\vec E(\vec r)\ldots,$ are distinct from the temporal Fourier transforms (phasor spectral densities)
$\vec J(\vec r,\omega), \vec E(\vec r,\omega),\ldots$:
\begin{align}
\label{eq-ph-phasor-transform}
    	\vec J(\vec r,t)
	&=\frac{1}{2}[\vec J(\vec r) e^{j\omega_0 t}+\vec J^*(\vec r) e^{-j\omega_0 t}]
	=\int_{-\infty}^{\infty}e^{j\omega t}\vec J(\vec r,\omega) d\omega
	\nonumber\\
	&=\int_{-\infty}^{\infty}e^{j\omega t}
	\left\{
	\frac{1}{2}[\vec J(\vec r)\delta(\omega-\omega_0)+\vec J^*(\vec r)\delta(\omega+\omega_0)]
	\right\}
	d\omega
\end{align}
where we use a Fourier transform convention consistent with \cite{1984-Haus,2011-Schachter}.
Equation (\ref{eq-ph-phasor-transform}) implies that for time-harmonic fields, the Fourier transforms can be expressed in terms of the corresponding phasors as
\begin{subequations}
\begin{align}
	\vec J(\vec r,\omega) &= \frac{1}{2}[\vec J(\vec r)\delta(\omega-\omega_0)+\vec J^*(\vec r)\delta(\omega+\omega_0)]
    \\
    	\vec E(\vec r,\omega) &= \frac{1}{2}[\vec E(\vec r)\delta(\omega-\omega_0)+\vec E^*(\vec r)\delta(\omega+\omega_0)], \ \text{etc.}
\end{align}
\end{subequations}
Note the difference in units: 
$\vec J(\vec r,\omega)\ \mathrm{[s \cdot (A/m^2)]}$, 
$\vec J(\vec r)\ \mathrm{[A/m^2]}$, etc.

Maxwell equations in the frequency domain (,,time-harmonic'')
\begin{subequations}
\label{eq-Maxwell}
\begin{gather}
    \nabla\times\vec E+j\omega\vec B=0\\
    \nabla\times\vec H-j\omega\vec D=\vec J\\
    \nabla\cdot\vec D=\rho\\
    \nabla\cdot\vec B=0
\end{gather}
\end{subequations}
are valid both for phasors and for temporal Fourier transforms, and this may lead to confusion about the meaning of symbols $\vec J, \vec E,\ldots$ in a particular context.
A frequency-domain solver in engineering-oriented software like Comsol expects a phasor expression for electric current $I(\vec r)$ [A] (or current density $\vec J(\vec r)$ [A/m$^2$]), and outputs
phasors $\vec E(\vec r)$ [V/m] and $\vec B(\vec r)$ [T]. The total energy radiated through a surface is
\begin{equation}
\label{eq-W-phasors}
    W=\int_{-\infty}^\infty \int_\text{surface}
    \Re\left[\frac{1}{2}\vec E(\vec r)\times \vec H^*(\vec r)\right]
    \cdot d\vec A\,dt
\end{equation}
where $\frac{1}{2}\vec E\times \vec H^*$ is the complex Poynting vector \cite{2011-Schachter} and
$\Re[\frac{1}{2}\vec E\times \vec H^*]$ is the time-averaged power flux density $\langle\vec{PFD}\rangle$ $[\mathrm{W/m^2}]$ (in Comsol it is called ``Power flow, time average, \texttt{Poav}''). Note that for strictly harmonic fields expression~(\ref{eq-W-phasors}) is infinite.

We can ``cheat'' the solver by entering a temporal Fourier transform for current
$I(\vec r,\omega)$ instead of a phasor $I(\vec r)$, then the solver 
will use the same equations (\ref{eq-Maxwell}) as for phasors to
calculate the Fourier transforms $\vec E(\vec r,\omega)$, $\vec B(\vec r,\omega),\ldots$ Now the expression $\Re[\frac{1}{2}\vec E\times \vec H^*]$ has a different interpretation and a different unit $[\mathrm{s^2 \cdot W/m^2}]$. As shown in Appendix \ref{section-derivation-W}, the total radiated energy is now equal to
\begin{equation}
\label{eq-W-transforms}
    W
    =\int_{0}^\infty 
    \underbrace{
    \int_\text{surface}
    4\cdot2\pi\cdot
    \Re\left[\frac{1}{2}\vec E(\vec r,\omega)\times \vec H^*(\vec r,\omega)\right]
    \cdot d\vec A
    }_{\displaystyle dW/d\omega}
    d\omega
\end{equation}
where $dW/d\omega$ is the radiated energy per unit frequency. Here we use  only positive frequencies to allow comparison with experimental results. The numerical factor $4\cdot 2\pi$ depends on which convention for Fourier transforms is used, here it is consistent with Equations~(\ref{eq-ph-phasor-transform}), (\ref{eq-current3}), (\ref{eq-current-transform}). To obtain $dW/d\omega$ from the solver's result, take the surface integral of $\langle\vec{PFD}\rangle$ (Comsol: \texttt{Poav}) and multiply by $4\cdot 2\pi$; the result is in $[\mathrm{J\cdot s}]$ (for phasors it would be $[\mathrm{J/s}]$). Depending on the software used, the ``cheated'' solver may signal wrong units. This can be resolved by multiplying the expression for current by an arbitrary frequency range $\Delta\omega$, for example by unit angular frequency
$\Delta\omega=1$~[1/s] and dividing $\langle\vec{PFD}\rangle$ by $(\Delta\omega)^2$.

The expression $\Re[\frac{1}{2}\vec E\times \vec H^*]$ is discussed in many electrodynamics texts for phasors $\vec E(\vec r)$, $\vec H(\vec r)$, but its interpretation for transforms $\vec E(\vec r,\omega )$, $\vec H(\vec r,\omega)$ and the formula (\ref{eq-W-transforms}) cannot easily be found in textbooks. To calculate power using a phasor-based frequency-domain code, one needs to compare directly equations (\ref{eq-W-phasors}) and (\ref{eq-W-transforms}). Formulas similar to (\ref{eq-W-transforms}) appear in some papers dealing with analytical models of S-P radiation \cite{1973-Lalor, 1984-Chuang-Kong,2018-Pan-Gover}, but may be not evident to users of numerical frequency-domain codes
(for example, this issue is not addressed in tutorials and manuals of Comsol \cite{Comsol}). Ref.~\cite{2019-Szczepkowicz} presents a heuristic argument for energy computation in a frequency domain solver, which however leads to results that are too small by a factor of 4. Adding to confusion, the expression $\Re\left[\frac{1}{2}\vec E(\vec r,\omega)\times \vec H^*(\vec r,\omega)\right]$ is sometimes called ``Poynting vector in the frequency domain'' \cite{1984-Chuang-Kong}, which can easily be misunderstood as the Fourier transform of the Poynting vector (these are distinct quantities with different units, see also Appendix \ref{section-derivation-W}). Another source of confusion is that energy density may be defined on either $\omega\in(-\infty,\infty)$ or $\omega\in(0,\infty)$ -- the two definitions differ by factor 2.
Incorrect numerical factors can also arise when rivaling Fourier transform conventions from different papers are confused. All these problems call for some verification of calculated radiation intensities. This issue will be addressed in Sect.~\ref{sect-verification-frank-tamm}.

\subsection{Moving point charge -- expression for current and Floquet-periodicity}

To calculate Smith-Purcell radiation or Cherenkov radiation, we consider
a point charge $(-e)$ moving in the $\hat z$ direction with constant velocity $v=\beta c$. For a particle at $z=0$ when $t=0$ the current density is
\begin{align}
    \vec J(\vec r,t)=(-e)v\delta(x)\delta(y)\delta(z-vt)\hat z
    =(-e)\delta(x)\delta(y)\delta(z/v-t)\hat z.
\end{align}
After integration  over the transverse coordinates $x,y$, we obtain
the expression for the current
\begin{align}
    I(z,t)&=(-e)\delta(z/v-t)= (-e)\frac{1}{2\pi}\int_{-\infty}^{\infty}e^{-j\omega (z/v-t)}d\omega
    \nonumber\\
    \label{eq-current3}
    &=\int_{-\infty}^{\infty}
    \left\{\frac{(-e)}{2\pi}e^{-j(\omega/v)z}\right\}
    e^{j\omega t} d\omega,
\end{align}
so the temporal Fourier transform of $I(z,t)$ is
\begin{equation}
    \label{eq-current-transform}
    I(z,\omega)=\frac{(-e)}{2\pi}e^{-j(\omega/v)z}.
\end{equation}
This expression is input to the solver as the ``edge current''. Note that the current is a function of the longitudinal coordinate and not all of the frequency-domain solvers will allow for this dependence on position.
The current is Floquet-periodic:
\begin{equation}
    \label{eq-current-periodic}
    I(z+a,\omega)=I(z,\omega)\exp(-j k_{\mathrm F}\, a),
\end{equation}
with the Floquet vector $k_{\mathrm F}=\omega/v$. For a uniform or a periodic medium, the same spatial periodicity in the frequency domain is followed by the fields:
\begin{align}
    \label{eq-E-periodicity}&\vec E(x,y,z+a,\omega)=\vec E(x,y,z,\omega) \exp({-j k_{\mathrm F}\, a}),\\
    \label{eq-H-periodicity}&\vec H(x,y,z+a,\omega)=\vec H(x,y,z,\omega) \exp({-j k_{\mathrm F}\, a}).
\end{align}
For Cherenkov radiation in a uniform medium $a$ is arbitrary. For a non-uniform medium the Floquet periodicity occurs if the refractive index is periodic in $z$ (eg.\ an infinite grating) -- in this case $a$ must be equal to the period. For non-periodic systems (eg.\ a finite grating) Eqs.~(\ref{eq-E-periodicity}--\ref{eq-H-periodicity}) do not hold.
However, they should hold in an approximate sense near the center of a finite structure with many periods. 

\subsection{Inifinte gratings -- geometry of the model}

In this paper we consider infinite gratings, periodic in the $z$ coordinate.
The described model is valid for arbitrary grating profile. We choose to focus on simple grating profiles: rectangular, and in Sect.~\ref{sect-triangular} -- triangular. As expected for numerical models, there is no additional difficulty if any other grating profile is considered, as long as the profile is well resolved by the calculation mesh (for 3D models computer memory limits may come into play).

\begin{figure}
\centering
\includegraphics{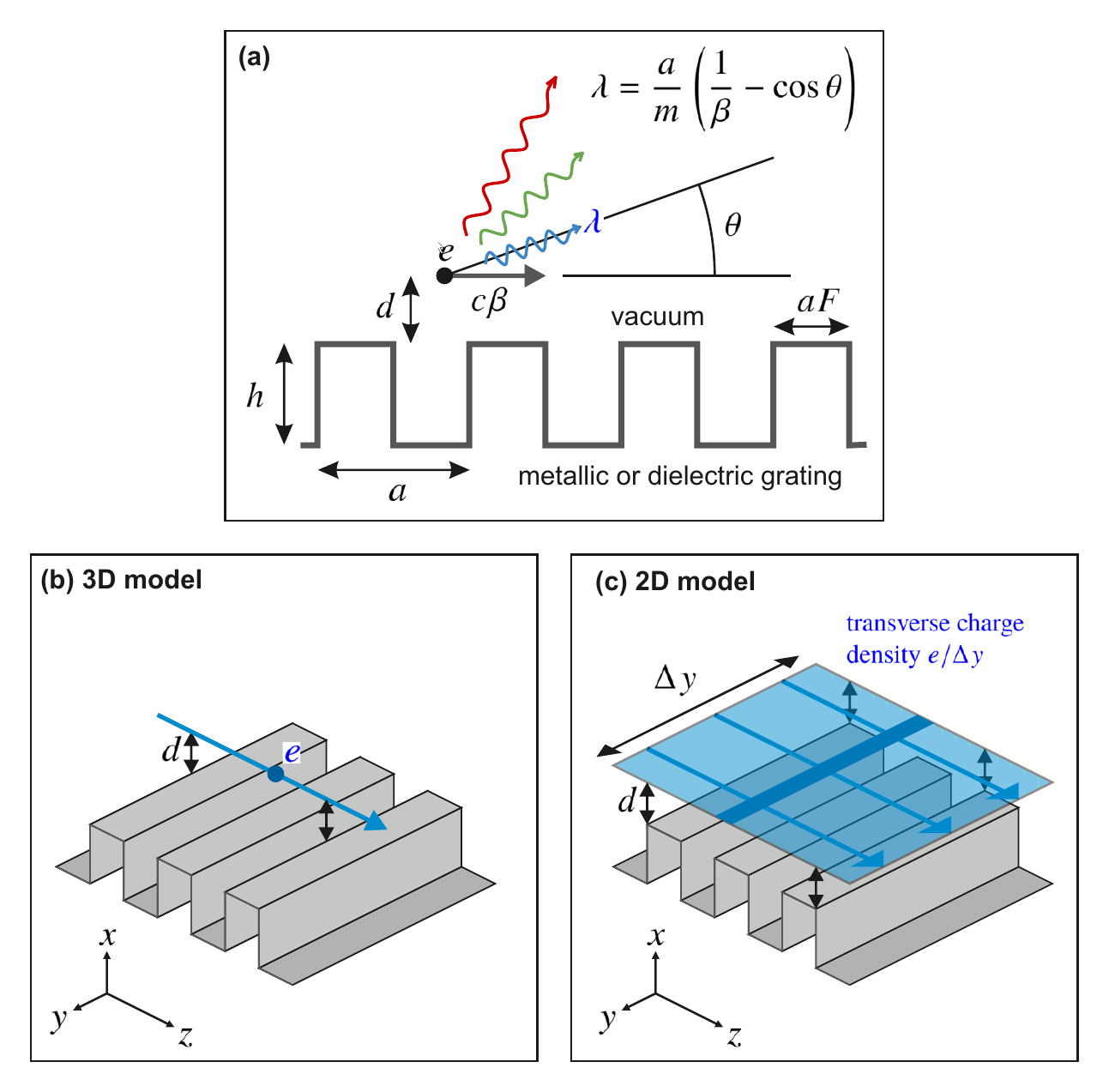}
\caption{\label{fig-grating}
(a) Basic grating and beam configuration assumed in this paper:
grating period $a=300$~nm, tooth height $h=200$~nm,
fill factor $F=0.5$,
$\beta=0.328$ corresponding to 30~keV electrons, impact parameter $d=100$~nm. We focus on the radiation in first spectral order $m=1$. For $\theta=90^{\circ}$ the radiation wavelength is $\lambda_{\perp}=914$~nm.
(b) A 3D model with a moving point charge; the grating is infinite in the $y$ and $z$ directions.
(c) A 2D model, invariant in the $y$ direction and infinite in the $z$ direction, with a moving line charge (flat beam pulse).}
\end{figure}

Figure \ref{fig-grating} shows the basic grating and beam configuration assumed in this paper. 
The electric charge moving in the $\hat z$ direction at a distance $d$ from a grating is the source of SP radiation. In the 3D model the source is a point charge $e$, and all radiation at a given frequency $\omega$ is collected. The 2D model corresponds to the source being a line charge with charge density $e/\Delta y$, and radiation collected from a corresponding strip of width $\Delta y$. In the latter case the resulting radiated energy depends on the arbitrary transverse length $\Delta y$ -- this issue will be addressed in Sect.~\ref{sect-comparison-3d-2d}.

\begin{figure}
\centering\includegraphics{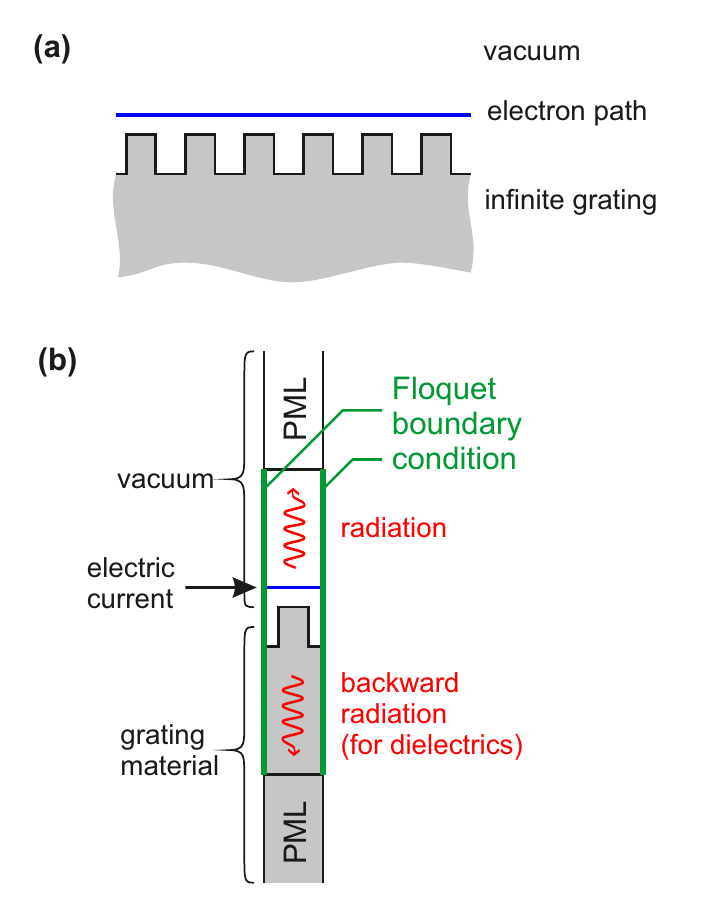}
\caption{\label{fig-2d-model} (a) A 2D infinite grating is modelled as a (b) 2D unit cell
with the Floquet boundary conditions.}
\end{figure}

\begin{figure}
\centering\includegraphics{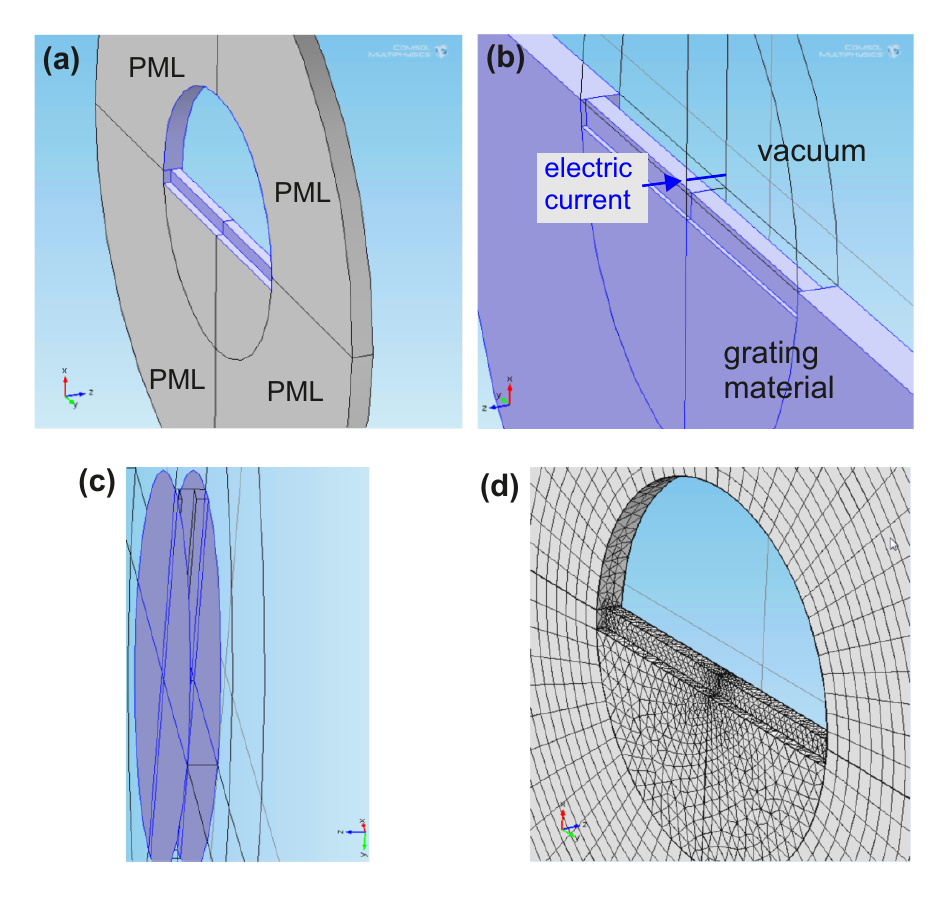}
\caption{\label{fig-3d-model}A 3D infinite grating is modelled as a 3D unit cell
with the Floquet boundary conditions.
(a) Perfectly matched layers (PMLs).
(b) Electron path.
(c) A pair of surfaces with Floquet boundary condition. 
(d) An example mesh used by a finite element method solver (Comsol).}
\end{figure}

Figures \ref{fig-2d-model} and \ref{fig-3d-model} show the details of the model.
For infinite gratings only one unit cell is needed in the calculation. 
The boundaries of the calculation domain in the $\pm\hat z$ directions are connected by the Floquet boundary condition (\ref{eq-E-periodicity}--\ref{eq-H-periodicity}). Perfectly Matched Layers (PMLs) attenuate radiation in the transverse directions.
The calculation domain consists of vacuum, characterized by relative permittivity = 1, and the grating region, characterized by arbitrary complex relative permittivity $\epsilon$. Note that ``top'' PMLs are vacuum, while the ``bottom'' PMLs are grating material, see Figs.~\ref{fig-2d-model}(b), \ref{fig-3d-model}(b). This means that we model thick gratings (this case occurs more often in the experiments). If the bottom PML is changed to vacuum, a thin grating is modelled, reflections from the back side of the grating occur, and guided mode resonances occur for certain frequencies (see Ref.~\cite{2016-Szczepkowicz} and references therein), with major impact on the radiation intensity. 
The 3D model is more realistic than a 2D model, but requires large RAM memory (10--20 GB); the computation time for a single frequency is of the order of minutes on a single CPU machine. 
The 2D model requires much less memory and the computation time is of the order of seconds. 
In the 3D simulation, radiation at a given frequency $\omega$ is collected from all transverse directions, by calculating the surface integral in Eq.~(\ref{eq-W-transforms}) over the inner boundary of the PML  (violet cylindrical strip in Fig.~\ref{fig-3d-model}(a)). In the 2D model, for a given frequency $\omega$, the integration is carried out over the planar inner PML boundary (Fig.~\ref{fig-2d-model}(b)).

We performed all calculations using the frequency-domain solver in the Comsol simulation software environment \cite{Comsol}, which utilizes the Finite Element Method (FEM). The FEM mesh is shown
in Fig.~\ref{fig-3d-model}(d). We tried various refinements of the mesh and obtained slight scatter of the computed results. On the basis of these trials we believe that the numerical values presented in this paper are accurate up to $\pm 20\%$, limited by the used hardware (20 GB RAM). Frequency-domain FEM method can also be applied to ultrarelativistic particles \cite{2020-Hausler}, but the mesh has to be refined along $z$ to account for the contraction of the electromagnetic field of the electron. We note one numerical peculiarity: in Comsol, for 3D models, the default iterative solver fails to find a solution within one hour, but if the direct solver is chosen, the solution is found within minutes.

\subsection{\label{sect-verification-frank-tamm}Verification of the model against the Frank-Tamm formula}

\begin{figure}
\centering\includegraphics{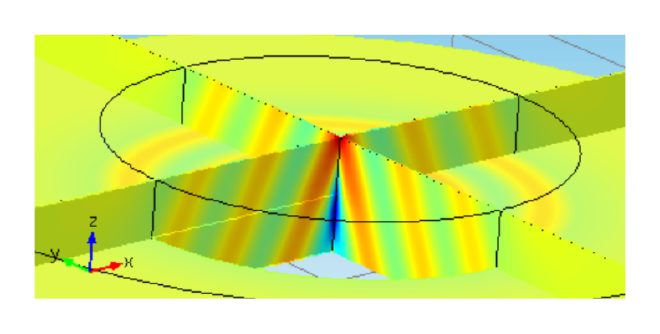}
\caption{\label{fig-cherenkov}Cherenkov radiation -- visualization of the Fourier transform of the electric field, $\Re[E_z(\vec r,\omega)]$, for $\omega=2\pi\cdot4\cdot10^{14}~\mathrm{s^{-1}}$, $\beta=0.5$, $n=5$.}
\end{figure}

At the end of Sect.~\ref{sect-phasors-vs-densities} we pointed out several pitfalls that may lead to incorrect multiplicative factors in calculations of SP radiation intensity. This doesn't matter if only relative intensity is needed or one aims at order-of-magnitude estimates, but if accurate results are needed, it is best to initially test the used model against some rigorous analytical result. We propose a novel approach to this problem. We take advantage of the fact that the numerical model described above is the same for SP and for Cherenkov radiation -- the only difference is in the distribution of relative permittivity in space, $\epsilon(\vec r)$.
So for a moment we change the relative permittivity of the vacuum region to $\epsilon$ and calculate the radiated energy in the uniform medium -- the Cherenkov radiation, shown in Fig.~\ref{fig-cherenkov}, 
and compare the radiation intensity with the Frank--Tamm formula, Eq.~(\ref{eq-Frank-Tamm}) in Appendix \ref{sect-derivation-Frank-Tamm}. The results are shown in Table \ref{table-cherenkov}.
\begin{table}
\centering
\begin{tabular}{|c|c|} 
 \hline
 Velocity and refractive index
 & $\left.\displaystyle\frac{\mathstrut d^2 W_\mathrm{num}}{\mathstrut dz\,d\omega} \middle/ \displaystyle\frac{\mathstrut d^2 W_\mathrm{analytical}}{\mathstrut dz\,d\omega}\right.$\rule[-3.2ex]{0em}{7.6ex} \\
 \hline\hline
 $\beta=0.33$, $n=3.6$ & 1.04 \\ 
 $\beta=0.5$, $n=3.6$ & 1.08 \\ 
 $\beta=0.5$, $n=5$ & 1.04 \\ 
 $\beta=0.3$, $n=2$ & no radiation \\ 
 \hline
\end{tabular}
\caption{\label{table-cherenkov}Cherenkov radiation -- verification of the numerical results against the Frank-Tamm formula for one frequency, $\omega=2\pi\cdot4\cdot10^{14}~\mathrm{s^{-1}}$. Here we compare the numerical value $d^2 W_\mathrm{num}/dz\,d\omega$ against the analytical value $d^2 W_\mathrm{analytical}/dz\,d\omega = (e^2/4\pi)\mu_0\omega(1-1/\beta^2n^2)$ (see Appendix~\ref{sect-derivation-Frank-Tamm}); the second column shows the ratio of these two values.}
\end{table}
The ratio of the numerical radiated energy $d^2 W_\mathrm{num} / dz\, d\omega$ to the analytical value is close to 1, within 10\%. The slight discrepancy can be reduced by refining the FEM mesh (within accessible computer memory). The parameters in the last row in the table do not fulfil the Cherenkov radiation condition (see Appendix~\ref{sect-derivation-Frank-Tamm}); in this case the calculated energy is 20 orders of magnitude lower, on the level of numerical noise of the calculation -- this is the expected result.
We also checked that the computed energy is linear in $\omega$. These results demonstrate that our model is exact, with no spurious multiplicative factors.

\section{\label{sect-comparison-materials}Comparison of radiation from metallic and dielectric gratings}

We assume grating and beam configuration described in Fig.~\ref{fig-grating}(a--b), a 3D model, and calculate the SP radiation emitted within the frequency range
$2\pi\cdot 325.5\mathrm{~THz}<\omega<2\pi\cdot 330.5\mathrm{~THz}$; this corresponds to radiation emitted perpendicular to the grating within 5~THz bandwidth, into angles $88.7^{\circ}<\theta<91.3^{\circ}$; the corresponding wavelength range is $907~\mathrm{nm}<\lambda<921~\mathrm{nm}$; see the SP formula in Fig.~\ref{fig-grating}(a).

To simulate a perfect conductor, we set the relative permittivity to $\epsilon = -10000+0i$ -- large negative real part and zero imaginary part. This is based on the observation that for optical frequencies good conductors have a large negative real part of $\epsilon$ 
(eg.
$\epsilon(\text{Cu}) =-36.8+1.36i$,
$\epsilon(\text{Au}) =-38.4+1.46i$ 
at $\lambda=914$~nm
\cite{2015-McPeak-Jayanti,refractiveindexinfo}%
),
while the imaginary part of $\epsilon$ describes energy dissipation and should vanish for a perfect conductor. This corresponds to large imaginary index of refraction $n=100i$, yielding almost instantaneous decay of the field inside the material (skin depth = $0.0016\lambda $ = 1.5 nm).
We verified numerically that for S--P radiation from metallic gratings in the considered frequency range, the bulk condition $\epsilon = -10000+0i$ yields the same result as the surface PEC condition (Perfect Electric Conductor, $\hat n \times \vec E = \vec 0$); this would not be valid in the sub-THz frequency range \cite{2010-Sukhikh-Naumenko}. While the boundary condition is computationally more efficient, we choose to use the bulk condition, so that exactly the same numerical model can be applied both to metallic and dielectric gratings –- only the grating’s relative permittivity $\epsilon$ is changed.

The resulting SP radiation for five materials is shown in Table~\ref{table-comparison-materials}.
\begin{table}
\centering
\begin{tabular}{|c|c|c|c|c|} 
 \hline
   \multicolumn{1}{|p{7em}|}{\centering Grating \\ material} &
   \multicolumn{2}{|p{10em}|}{\centering Relative permittivity \\ ($\lambda=914$~nm)} &
   \multicolumn{1}{|p{7em}|}{\centering Radiated energy \\ $W$ } &
   \multicolumn{1}{|p{8em}|}{\centering Energy radiated into \\ the grating $W_\mathrm{in}$}
   \\
   \cline{2-3}
   &
   \multicolumn{1}{|p{4em}|}{\centering Real part \\ $\epsilon_1$} &
   \multicolumn{1}{|p{4em}|}{\centering Imaginary \\ part $\epsilon_2$} & &
   \\
 \hline\hline
  Copper & $-36.85$ & 1.361 & $4.4\cdot 10^{-25}$ J & 0 \\
  Gold   & $-38.36$ & 1.462 & $4.4\cdot 10^{-25}$ J & 0 \\
  Perfect conductor & $-10000$ & 0 & $3.1\cdot 10^{-25}$ J & 0 \\
  Fused silica & 2.107 & 0 & $4.8\cdot 10^{-27}$ J & $8.8\cdot10^{-27}$ J \\
  Silicon & 13.32 & 0.03099 & $3.0\cdot 10^{-26}$ J & $6.8\cdot 10^{-26}$ J \\
 \hline
\end{tabular}
\caption{\label{table-comparison-materials}SP radiation from gratings of different materials, emitted perpendicular to the grating within the frequency range
$2\pi\cdot 325.5\mathrm{~THz}<\omega<2\pi\cdot 330.5\mathrm{~THz}$, corresponding to angular range $88.7^{\circ}<\theta<91.3^{\circ}$,
per electron per grating period, for the grating geometry and beam parameters from Fig.~\ref{fig-grating}(a--b). Relative permittivity is taken from Refs. \cite{refractiveindexinfo, 1965-Malitson,2015-McPeak-Jayanti, 2015-Schinke-Peest}.}
\end{table}
Metallic gratings outperform dielectric gratings by 1--2 orders of magnitude in terms of radiation intensity under the conditions studied (simple rectangular grating, 30 keV electrons, radiation wavelength $\sim1$~um). This confirms expectations from comparison of theoretical upper bounds for radiation from Ref.~\cite{2019-RoquesCarmes-Kooi}
(plot of ``material factor'' in Supplementary Fig.~10).
An interesting observation is that for dielectric gratings twice more energy is radiated inside the grating than emitted as external SP radiation. It is an interesting question whether the energy radiated inside could be reflected using eg.\ a vacuum/dielectric Bragg mirror (DBR) \cite{2019-Yousefi-Schonenberger}.

Data in Table~\ref{table-comparison-materials} is a direct comparison of SP radiation from metallic and dielectric gratings, however it is only for one set of beam and grating parameters. In future work the study should be extended to other electron energies, grating configurations and radiation angles. In particular, it would be interesting to check the prediction from Ref.~\cite{2019-RoquesCarmes-Kooi} that silicon may outperform metals for very low energy electrons (below 10~eV).

\section{\label{sect-oscillations}Energy oscillations with grating tooth height -- grating resonances}

\begin{figure}
\centering\includegraphics{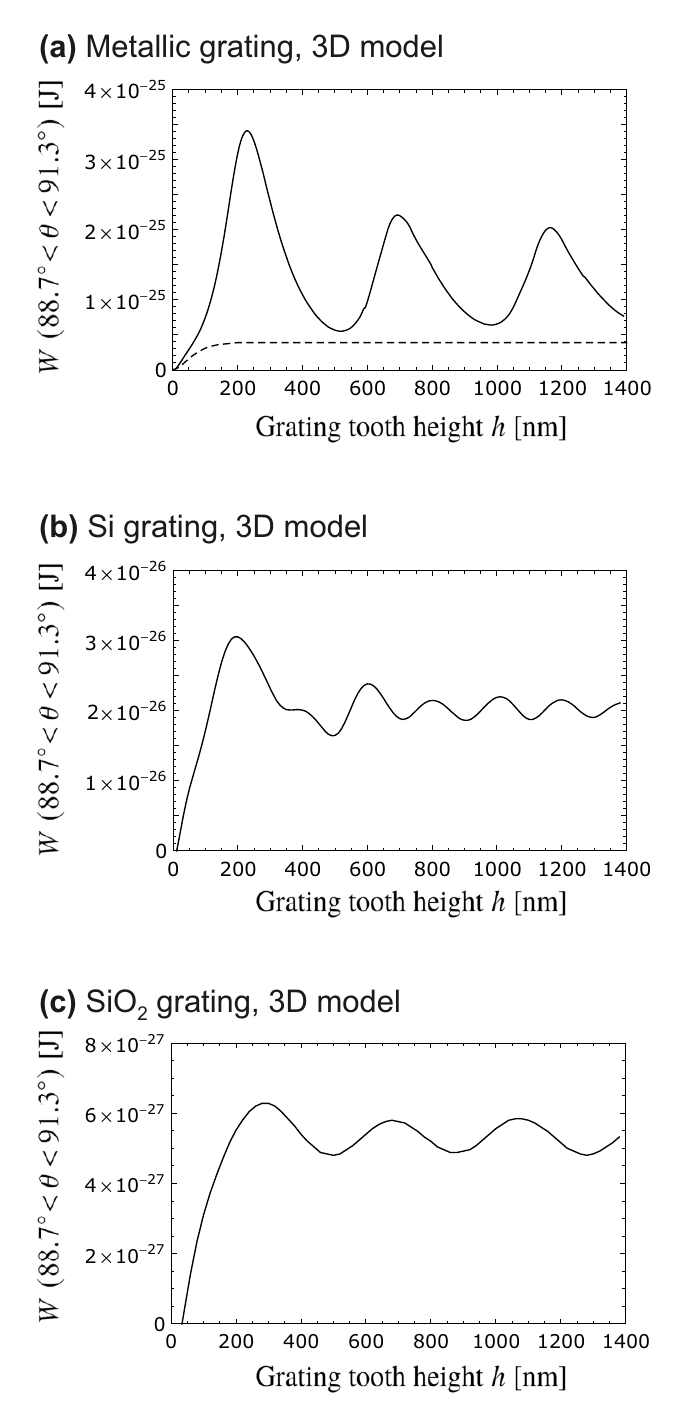}
\caption{\label{fig-oscillations-comparison-of-materials}SP radiation from gratings of different materials, 
in the angular range $88.7^{\circ}<\theta<91.3^{\circ}$,
per electron per grating period, for grating geometry shown in Fig.~\ref{fig-grating}(a--b) (3D model).
Solid line -- numerical calculation in the frequency domain. For the metallic grating the result is compared with an analytical calculation based on Ref.~\cite{2005-Brownell-Doucas} -- dashed line.
}
\end{figure}

Results of the previous section were for a grating with a fixed tooth height. Does the intensity of SP radiation change if the tooth height is changed? The answer is shown in Fig.~\ref{fig-oscillations-comparison-of-materials}.
SP radiation intensity $W$ oscillates with increasing tooth height $h$ (the impact parameter $d$ is kept constant, see Fig.~\ref{fig-grating}). The effect is known for metallic gratings, 
it has been experimentally observed in the millimeter-wave spectral region using a special metallic grating setup with variable tooth height \cite{1998-Shibata-Hasebe}. The effect has also been shown for metallic gratings in an analytical calculation based on van den Berg's model \cite{1998-Shibata-Hasebe} and in numerical time-domain models \cite{1998-Shibata-Hasebe, 2014-Liu-Xu, 2015-Liu-Li,2015-Lekomtsev-Karataev}. Here we confirm oscillations of $W(h)$ for metallic gratings in a frequency-domain simulation; in addition, we demonstrate that the oscillations also occur for dielectric gratings, see Fig.~\ref{fig-oscillations-comparison-of-materials}(b--c). An interesting result is that for all materials, the optimum tooth height is close to one quarter of the radiation wavelength in the perpendicular direction.

\begin{figure}
\centering\includegraphics{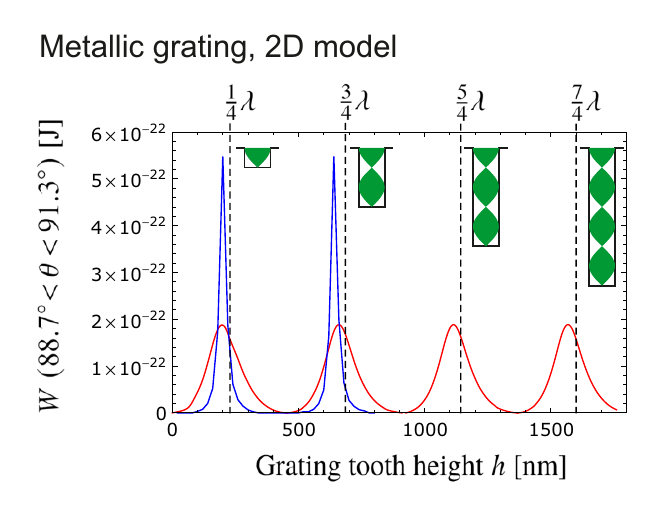}
\caption{\label{fig-oscillations-metallic-2d}SP radiation from gratings of different tooth heights $h$, 
in the angular range $88.7^{\circ}<\theta<91.3^{\circ}$,
per electron per grating period, for grating geometry shown in Fig.~\ref{fig-grating}(a,c) (2D model) with $\Delta y=1$~nm. Grating fill factor $F=0.5$ (red) and $F=0.9$ (blue).}
\end{figure}

The $W(h)$ oscillations are of resonant nature. Similar effects for the more general phenomenon of diffraction radiation \cite{2011-Potylitsyn-Ryazanov} have been predicted for the motion of charged particles near an open metallic or dielectric resonator \cite{1981-Nosich,2019-Yevtushenko-Dukhopelnykov}  
or an array of metallic resonators \cite{1977-Veliev-Nosich};
enhanced radiation occurs for frequencies close to one of the resonant frequencies of the cavity. In case of the oscillatory effect shown in Fig.~\ref{fig-oscillations-comparison-of-materials}, the cavity is formed between adjacent grating teeth. The essence of the effect is already captured by a 2D model, see Fig.~\ref{fig-oscillations-metallic-2d}.
The green insets show the analogy with the elementary one-dimensional theory of clarinets or organ pipes, where the resonant lengths of a cavity which is open at one end, for a fixed wavelength $\lambda$, are equal to $\frac{1}{4}\lambda+m\cdot\frac{1}{2}\lambda$ (vertical dashed lines in Fig.~\ref{fig-oscillations-metallic-2d}; see also Ref.~\cite{2014-Liu-Xu}). This simple one-dimensional reasoning explains the essence of the effect, but does not explain the slight shift of the calculated radiation maxima $W(h)$ towards smaller $h$.
Note that in the acoustic case this model is also approximate.
The simple model does not work in case of dielectric cavities
-- the period of $W(h)$ oscillations in Fig.~\ref{fig-oscillations-comparison-of-materials}(b--c) has no obvious relation to the vacuum wavelength or the wavelength inside the dielectric; maybe one could deduce the effective wavelength on the grounds of effective mode index theory.

The intensity oscillations for a metallic grating can alternatively be explained by considering the groove cavity as a transmission line which is short-circuited at one end (bottom of the groove; impedance = 0) and open-circuited at the other end (top of the groove). If in the groove only the TEM mode is considered (transmission line approximation), the first resonance will occur when $h=\lambda/4$ (high impedance at the top end of the line). However, in practice many other modes must be used in order to satisfy the boundary conditions. Due to the presence of these evanescent modes electromagnetic energy is stored near the edge (top) of the groove. Effectively, this stored energy is manifested in a shift in frequency (see Fig.~\ref{fig-oscillations-metallic-2d}), since in general near an edge the magnetic energy does not equal the electric energy – as is the case in an ideal cavity or plane waves. This effect is well known in waveguide theory and it is usually described in terms of length: the electrical length differs from the geometrical length.

The van den Berg's model cannot be applied to dielectrics, but it would be interesting to compare it with oscillations shown in Fig.~\ref{fig-oscillations-comparison-of-materials}(a) for metallic grating, similarly as was done in Ref.~\cite{1998-Shibata-Hasebe}. However, van den Berg's model is not easy to apply, so instead we compare our numerical results with the analytical results of a surface current model, using expressions from Ref.~\cite{2005-Brownell-Doucas} (dashed line in Fig.~\ref{fig-oscillations-comparison-of-materials}(a)). An order of magnitude agreement is found, but the surface current model does not predict oscillations of $W(h)$. The explanation can be found in the original paper by Brownell, Walsh and Doucas:
``in deep tooth profiles several facets may form a
cavity and limit the field modes when the wavelength is
comparable to or longer than the cavity dimensions. The
model described here is best suited for shallow gratings
where cavity behavior is negligible but can be applied to
deep profiles if the energy is sufficiently high so that the
wavelength is much smaller than any cavity'' \cite{1998-Brownell-Walsh}.

\section{\label{sect-comparison-3d-2d}Comparison of 3D and 2D results for the Smith-Purcell radiation}

As was shown in the previous section, certain effects in the SP radiation can be captured well within a 2D model, which is much simpler to construct than a 3D model. What about radiation intensity: can we estimate it from a 2D model? Comparison of the numbers in Figs.~\ref{fig-oscillations-comparison-of-materials}(a) and \ref{fig-oscillations-metallic-2d} reveals a difference in energy by 3 orders of magnitude. One should not expect agreement, because the two physical situations are different: single electron vs.\ an infinite line charge with linear density $\rho=e/\Delta y$. The latter contains an arbitrary parameter $\Delta y$, and the final calculated energy $W$ is inversely proportional to this parameter ($W \propto \rho^2\Delta y = (e/\Delta y)^2\Delta y = e^2/\Delta y$ -- we are grateful to Urs H\"{a}usler for pointing this out).
For Fig.~\ref{fig-oscillations-metallic-2d} the parameter $\Delta y$ was arbitrarily chosen to equal 1~nm (radiation is generated by a line of charge with a transverse charge density of $e/1$~nm; radiation $W$ is collected from a longitudinal strip of the grating of width 1 nm), a length not connected with the characteristic length scales of the considered grating/beam configuration. It appears reasonable to replace $\Delta y=1$~nm with $\Delta y=\lambda_\perp$, which is one of the characteristic lengths of the system. The result is shown in Table~\ref{table-comparison-3d-2d}.
\begin{table}
\centering
\begin{tabular}{|c|c|} 
 \hline
 Geometry of the model & Radiated energy $W$ \\
 \hline\hline
 3D model & $3.12\cdot10^{-25}$ J \\
 2D model, $\Delta y=\lambda_\perp=914$ nm & $2.02\cdot10^{-25}$ J \\
 2D model, $\Delta y=1$ nm & $1.85\cdot10^{-22}$ J\\
 \hline
\end{tabular}
\caption{\label{table-comparison-3d-2d}Comparison of 3D and 2D SP radiation in the angular range $88.7^{\circ}<\theta<91.3^{\circ}$ for a metallic grating with beam/grating configuration of Fig.~\ref{fig-grating}.}
\end{table}
The 2D result now predicts the order of magnitude of the 3D result. It remains an open question whether this trick would work equally well for other beam/grating configurations.

\section{\label{sect-triangular}Metallic triangular grating (3D)}

\begin{figure}[hbp]
\centering\includegraphics{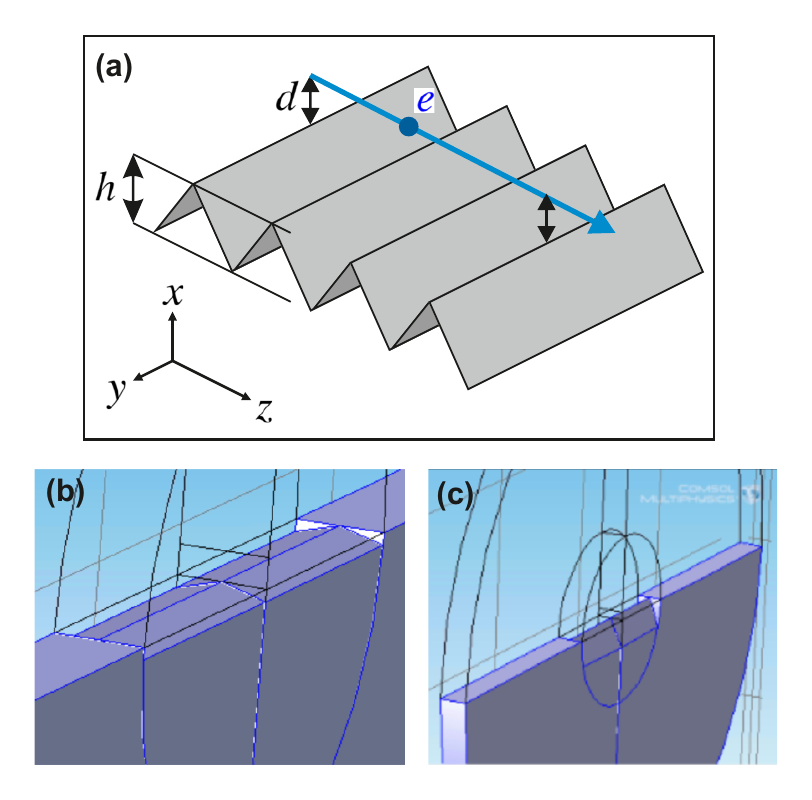}
\caption{\label{fig-3d-model-triangular}(a) An infinite triangular grating, 3D model. All parameters are the same as in Fig.~\ref{fig-grating}, but $h$ is varied.
(b) The unit cell for $h=40$~nm and (c) $h=320$~nm. In all cases the grating period and the impact parameter are $a=300$~nm, $d=100$~nm.} 
\end{figure}

In previous sections we considered rectangular gratings. In this section we briefly consider a 3D model of a grating with a triangular profile shown in Fig.~\ref{fig-3d-model-triangular}. 
This is motivated by the expectation that the surface current model of Ref.~\cite{2005-Brownell-Doucas} should work well for such a configuration, provided that the grating is shallow, $h\ll a$, so that resonant cavities do not form. We decided to test this hypothesis. In Table~\ref{table-triangular} we compare our 3D results with the analytical result based on formulas from Ref.~\cite{2005-Brownell-Doucas}.
\begin{table}
\centering
\begin{tabular}{|c|c|c|c|} 
 \hline
 Grating tooth height $h$ & $W$, numerical model & $W'$, analytical model & Ratio $W'/W$  \\
 \hline\hline
  \phantom040 nm &  $1.42\cdot10^{-27}$ J & $5.7\cdot10^{-27}$ J & 4.0\\ 
  \phantom080 nm &  $3.18\cdot10^{-27}$ J & $1.1\cdot10^{-26}$ J & 3.5\\ 
  160 nm & $4.96\cdot10^{-27}$ J & $1.2\cdot10^{-26}$ J & 2.4\\ 
  320 nm & $6.28\cdot10^{-27}$ J & $6.7\cdot10^{-27}$ J & 1.1\\ 
 \hline
\end{tabular}
\caption{\label{table-triangular}SP radiation from metallic triangular gratings, in the angular range $88.7^{\circ}<\theta<91.3^{\circ}$, per electron per grating period, for grating geometry shown in Fig.~\ref{fig-3d-model-triangular}. We compare our numerical results with the analytical model of Ref.~\cite{2005-Brownell-Doucas}}
\end{table}
Similarly as in Fig.~\ref{fig-oscillations-comparison-of-materials}(a), we obtain an order of magnitude agreement. However, contrary to our expectations, the agreement gets worse as the tooth height is decreased. We believe that this points either to the inexactness of surface current model even for shallow gratings, or to some multiplicative factor issue as discussed at the end of Sect.~\ref{sect-phasors-vs-densities}. This remains an open question, we have only checked that it is not the frequency range issue $(-\infty,\infty)$ vs.\ $(0,\infty)$, because Eqs~(14.60) and (14.70) in Ref.~\cite{1998-Jackson}, which are the starting point in Ref.~\cite{1998-Brownell-Walsh}, are for $\omega\in(0,\infty)$, same as assumed in our calculations.

\section{\label{sect-summary}Summary and conclusions}

We constructed a numerical frequency-domain model useful for quick calculations of Smith-Purcell radiation intensity from a single particle, verified its accuracy, and discussed some concrete applications. We were somewhat conservative in the choice of grating geometry (rectangular and triangular gratings of uniform material). This was a deliberate choice which facilitated comparison with the older literature. The paper in large part deals with the method, but also presents new results regarding comparison of SP radiation from metallic and dielectric gratings; the possible numerous other applications are left for future work. The main results of the paper can be summarized as follows:
\begin{enumerate}
    \item A frequency-domain numerical model offers quick calculations of SP radiation intensity (seconds for 2D models, minutes for 3D models, for one frequency, on a single CPU machine).
    
    \item While it is relatively easy to calculate SP radiation in arbitrary units, obtaining an accurate result in concrete units with no spurious multiplicative constants requires careful differentiation between phasors (quantities expected by the numerical solver) and phasor densities (temporal Fourier transforms). After this issue is properly taken care of, the model is simple, easy to implement, and its accuracy is limited only by the spatial mesh density, possibly limited by the amount of RAM computer memory accessible to the user.
    
    \item The accuracy of the numerical model can be conveniently checked against the analytical result: the Frank-Tamm formula.
    
    \item For 30 keV electrons dielectric gratings are less efficient in terms of intensity of SP radiation than their metallic counterpart, by an order of magnitude for silicon, and two orders of magnitude for silica. 
    
    \item The described numerical model captures resonant effects in the grating which are not accounted for in some analytical models. In particular, both for metallic and dielectric gratings, SP radiation intensity oscillates with grating tooth height. Both for metals and dielectrics, the optimum tooth height for maximal SP radiation perpendicular to the grating is close to a quarter of the radiation wavelength.
    
    \item For dielectric gratings, more SP radiation enters the grating bulk than is radiated outward into the vacuum.

\end{enumerate}

\section*{Funding}

This work was supported by the Gordon and Betty Moore Foundation
(grant no.\ GBMF4744) and the U.S. Department of Energy,
Office of Science (grant nos.\ DE-AC02-76SF00515 and
DE-SC0009914). LS was supported by the Israel Science Foundation.

\section*{Acknowledgments}
We are grateful to  Avi Gover for pointing us to the older literature on the Smith-Purcell effect and for helpful comments; to  Yen-Chieh Huang, Koby Scheuer, Naama Cohen and Urs H\"{a}usler for inspiring discussions.

A.S. is grateful to the Wroclaw Centre for Networking and Supercomputing for granting access to the Platon U3 computing infrastructure.

\section*{Disclosures}

The authors declare no conflicts of interest.

\appendix

\section{\label{section-derivation-W}Energy spectral density -- derivation of Eq.~(\ref{eq-W-transforms})}

We will show that
\begin{equation}
\label{eq-Parseval}
    \int_{-\infty}^\infty 
  \vec E(\vec r,t)\times \vec H(\vec r,t)\,dt
    =\int_{0}^\infty 
    4\cdot2\pi\cdot
    \Re\left[\frac{1}{2}\vec E(\vec r,\omega)\times \vec H^*(\vec r,\omega)\right]\,d\omega
\end{equation}
which is equivalent to Eq.~(\ref{eq-W-transforms}).
This formula is a version of the Parseval's theorem useful for frequency-domain electromagnetic calculations; it shows the 
connection between the energy distribution in time and energy distribution in frequency. All equations are consistent with the Fourier transform convention of Eq.~(\ref{eq-ph-phasor-transform}).

First we determine the Fourier transform of the Poynting vector.
\begin{align}
    \vec S(\vec r,t) &= \vec E(\vec r,t) \times \vec H(\vec r,t)
    = \int_{-\infty}^{\infty}\vec E(\vec r,\omega')e^{j \omega' t}d\omega' \times \int_{-\infty}^{\infty}\vec H(\vec r,\omega'')e^{j \omega'' t}d\omega'' \\
    &= \int_{-\infty}^{\infty} \int_{-\infty}^{\infty}
    e^{j(\omega'+\omega'')t} \vec E(\vec r,\omega') \times \vec H(\vec r,\omega'')\, d\omega'' d\omega'
\end{align}
After the change of variable in the inner integral,  $\omega=\omega'+\omega'', d\omega=d\omega''$, we obtain
\begin{align}
    \vec S(\vec r,t) &= 
    \int_{-\infty}^{\infty} \int_{-\infty}^{\infty} e^{j\omega t} \vec E(\vec r,\omega') \times \vec H(\vec r,\omega-\omega') \,d\omega\, d\omega' \\
    \label{eq-S-transform}
    &= \int_{-\infty}^{\infty} e^{j\omega t}
    \underbrace{
    \int_{-\infty}^{\infty} \vec E(\vec r,\omega') \times \vec H(\vec r,\omega-\omega') \,d\omega'
    }_{\displaystyle\vec S(\vec r,\omega)} d\omega
\end{align}
This shows that the Fourier transform of the Poynting vector $\vec S(\vec r,\omega)$ is the convolution of transforms of the fields (convolution theorem).

From the definition of the Poynting vector and Eq. (\ref{eq-S-transform}), the total energy radiated through a surface is
\begin{align}
 W &= \int_{-\infty}^{\infty} \int_{\text{surface}} \vec S(\vec r,t) \cdot d\vec A \,dt \\
 &= \int_{-\infty}^{\infty} \int_{\text{surface}} 
 \int_{-\infty}^{\infty} e^{j\omega t}
    \int_{-\infty}^{\infty} \vec E(\vec r,\omega') \times \vec H(\vec r,\omega-\omega') \,d\omega' d\omega
 \cdot d\vec A \,dt
\end{align}
Similarly as in Ref.~\cite{1998-Jackson}, chapter~14.5, we simplify the formula by noting a representation of the delta function,
$\int_{-\infty}^{\infty}e^{j \omega t}dt=2\pi \delta(\omega)$, and noting that $\vec H(\vec r,-\omega')=\vec H^*(\vec r,\omega')$ (a property of transforms of real functions):
\begin{align}
 W &= \int_{\text{surface}} 2\pi 
 \int_{-\infty}^{\infty} \vec E(\vec r,\omega') \times \vec H^*(\vec r,\omega') \,d\omega'
 \cdot d\vec A
\end{align}
In experimental investigations, a useful, measurable quantity is the distribution of energy in positive frequencies, so we reduce
the domain of integration over frequencies:
\begin{align}
 W &= \int_{\text{surface}} 2\pi 
 \int_{0}^{\infty} 
 \left[
 \vec E(\vec r,\omega') \times \vec H^*(\vec r,\omega') +
 \vec E^*(\vec r,\omega') \times \vec H(\vec r,\omega')
 \right] \,d\omega'  \cdot d\vec A \\
 &= \int_{\text{surface}} 4\cdot 2\pi 
 \int_{0}^{\infty} 
 \Re\left[ \frac{1}{2}
 \vec E(\vec r,\omega') \times \vec H^*(\vec r,\omega') 
 \right] \,d\omega'  \cdot d\vec A
\end{align}
The factor $\frac{1}{2}$ is for easy comparison with the energy formula for phasors (\ref{eq-W-phasors}).

\section{\label{sect-derivation-Frank-Tamm}Derivation of the Frank-Tamm formula from the potentials}

For additional verification of Eq.~(\ref{eq-W-transforms}), we will use it to derive the Frank--Tamm formula.
The potentials for a point charge $(-e)$ moving with velocity $\beta c$ in a uniform medium characterized by the refractive index $n$, can be expressed as
\begin{align}
    \label{eq-pot-A}&A_r=0, A_\phi=0, A_z(r,z,\omega)=(-e)\frac{\mu_0}{(2\pi)^2} {\mathrm K}_0\left(j\frac{\omega}{c}r\sqrt{n^2-\beta^{-2}}\right)\exp\left(-j\frac{\omega}{\beta c}z\right),\\
    \label{eq-pot-phi}&\Phi(r,z,\omega) = \frac{c}{n^2 \beta}A_z(r,z,\omega),
\end{align}
see Ref.~\cite{2011-Schachter}, Eqs~(2.1.36--38), (2.4.20).
${\mathrm K}_0$ is a modified Bessel function of the second kind. 
In subsequent calculations, to determine the energy flow in the far field, we use the first term of the asymptotic expansion of ${\mathrm K}_0$:
\begin{equation}
    \label{eq-K0-expansion}
    {\mathrm K}_0(j\xi) \simeq\sqrt{\frac{\pi}{2\xi}}e^{-j\pi/4}e^{-j\xi}\quad \mathrm{for\ } \xi\gg1 
\end{equation}
Consider a cylindrical surface of radius $r$ and length $\Delta z$ surrounding the electron trajectory. According to Eq.~(\ref{eq-W-transforms}), the energy radiated through this surface is
\begin{equation}
    \label{eq-W-cyl}
    \Delta W=\int_0^{\infty}\left\{ 4\cdot 2\pi\cdot(2\pi r \Delta z)\,\hat r \cdot \Re\left[\frac{1}{2}\vec E(\vec r,\omega)\times \frac{1}{\mu_0}\vec B^*(\vec r,\omega)\right]\right\}d\omega
\end{equation}
The fields are determined from the potentials in the usual manner -- the equations for transforms are the same as for phasors:
\begin{align}
    &\vec E(\vec r,\omega) = -j\omega \vec A(\vec r,\omega) - \nabla\Phi(\vec r,\omega)=-j\omega A_z\hat z-\frac{c}{n^2\beta}\left(\hat r\frac{\partial A_z}{\partial r}+\hat z\frac{\partial A_z}{\partial z}\right)\\
    &\vec B^*(\vec r,\omega) = \nabla \times \vec A^*(\vec r,\omega) = -\hat\phi\frac{\partial A_z^*}{\partial r}
\end{align}
Here the scalar potential was eliminated using Eq.~(\ref{eq-pot-phi}), and the fields are all expressed as a function of the longitudinal component of the vector potential. Derivatives of $A_z$ are computed using Eqs. (\ref{eq-pot-A}) and (\ref{eq-K0-expansion}). We obtain
\begin{align}
    \hat r\cdot [\vec E(\vec r,\omega)\times\mu_0^{-1}\vec B^*(\vec r,\omega)]=
    \mu_0^{-1}\left[ -j\omega + \frac{c}{n^2\beta}\left(j\frac{\omega}{\beta c}\right) \right] A_z \left[ -\frac{1}{2r}+j\frac{\omega}{c}\sqrt{n^2-\beta^{-2}} \right] A_z^*
\end{align}
The real part of this expression depends on whether the Cherenkov radiation condition is fulfilled ($\beta c>c/n$):
\begin{equation}
    \hat r\cdot \Re[\vec E(\vec r,\omega)\times\mu_0^{-1}\vec B^*(\vec r,\omega)]=
    \left\{
    \begin{array}{ll}
      \displaystyle \omega\left(1-\frac{1}{n^2\beta^2}\right)\frac{\mu_0 e^2}{32\pi^3 r} & \mathrm{for~~} n>1/\beta\\
      \displaystyle 0 & \mathrm{for~~} n<1/\beta \\
    \end{array} 
    \right.
\end{equation}
Insertion of this expression into Eq.~(\ref{eq-W-cyl}) yields the energy radiated by the charge per unit travelled length -- the Frank--Tamm formula \cite{1937-Frank-Tamm,1998-Jackson,2004-Shiozawa}:
\begin{equation}
    \label{eq-Frank-Tamm}
    \Delta W/\Delta z=
    \int_0^\infty
    \left\{
    \begin{array}{ll}
      \displaystyle e^2\frac{\mu_0}{4\pi}\omega\left(1-\frac{1}{n^2\beta^2}\right) & \mathrm{for~~} n>1/\beta\\
      \displaystyle 0 & \mathrm{for~~} n<1/\beta \\
    \end{array} 
    \right\} d\omega.
\end{equation}


\bibliography{1bibliography}

\begin{thebibliography}{10}
\newcommand{\enquote}[1]{``#1''}

\bibitem{1953-Smith-Purcell}
S.~J. Smith and E.~M. Purcell, \enquote{Visible light from localized surface
  charges moving across a grating,} {\protect\JournalTitle{Physical Review}}
  \textbf{92}, 1069--1069 (1953).

\bibitem{1969-Rusin-Bogomolov}
F.~Rusin and G.~Bogomolov, \enquote{Orotron{\textemdash}an electronic
  oscillator with an open resonator and reflecting grating,}
  {\protect\JournalTitle{Proceedings of the {IEEE}}} \textbf{57}, 720--722
  (1969).

\bibitem{2004-Grishin-Fuchs}
Y.~A. Grishin, M.~R. Fuchs, A.~Schnegg, A.~A. Dubinskii, B.~S. Dumesh, F.~S.
  Rusin, V.~L. Bratman, and K.~M\"{o}bius, \enquote{Pulsed
  orotron{\textemdash}a new microwave source for submillimeter pulse high-field
  electron paramagnetic resonance spectroscopy,} {\protect\JournalTitle{Review
  of Scientific Instruments}} \textbf{75}, 2926--2936 (2004).

\bibitem{2014-So-MacDonald}
J.-K. So, K.~F. MacDonald, and N.~I. Zheludev, \enquote{Fiber optic probe of
  free electron evanescent fields in the optical frequency range,}
  {\protect\JournalTitle{Applied Physics Letters}} \textbf{104}, 201101 (2014).

\bibitem{2019-Ye-Liu}
Y.~Ye, F.~Liu, M.~Wang, L.~Tai, K.~Cui, X.~Feng, W.~Zhang, and Y.~Huang,
  \enquote{Deep-ultraviolet {S}mith{\textendash}{P}urcell radiation,}
  {\protect\JournalTitle{Optica}} \textbf{6}, 592 (2019).

\bibitem{2019-RoquesCarmes-Kooi}
C.~Roques-Carmes, S.~E. Kooi, Y.~Yang, A.~Massuda, P.~D. Keathley, A.~Zaidi,
  Y.~Yang, J.~D. Joannopoulos, K.~K. Berggren, I.~Kaminer, and
  M.~Solja{\v{c}}i{\'{c}}, \enquote{Towards integrated tunable all-silicon
  free-electron light sources,} {\protect\JournalTitle{Nature Communications}}
  \textbf{10} (2019).

\bibitem{1989-Fernow}
R.~C. Fernow, \enquote{Design of a grating for studying {Smith-Purcell}
  radiation and electron acceleration,} in \emph{{AIP} Conference Proceedings,}
   ({AIP}, 1989).

\bibitem{2001-Doucas-Kimmitt}
G.~Doucas, M.~Kimmitt, J.~Brownell, S.~Trotz, and J.~Walsh, \enquote{A new type
  of high-resolution position sensor for ultra-relativistic beams,}
  {\protect\JournalTitle{Nuclear Instruments and Methods in Physics Research
  Section A: Accelerators, Spectrometers, Detectors and Associated Equipment}}
  \textbf{474}, 10--18 (2001).

\bibitem{2012-Soong-Byer}
K.~Soong and R.~L. Byer, \enquote{Design of a subnanometer resolution beam
  position monitor for dielectric laser accelerators,}
  {\protect\JournalTitle{Optics Letters}} \textbf{37}, 975 (2012).

\bibitem{1997-Lampel}
M.~Lampel, \enquote{Coherent {Smith-Purcell} radiation as a pulse length
  diagnostic,} {\protect\JournalTitle{Nuclear Instruments and Methods in
  Physics Research Section A: Accelerators, Spectrometers, Detectors and
  Associated Equipment}} \textbf{385}, 19--25 (1997).

\bibitem{1997-Nguyen}
D.~C. Nguyen, \enquote{Coherent smith-purcell radiation as a diagnostic for
  subpicosecond electron bunch length,} {\protect\JournalTitle{Nuclear
  Instruments and Methods in Physics Research Section A: Accelerators,
  Spectrometers, Detectors and Associated Equipment}} \textbf{393}, 514--518
  (1997).

\bibitem{2002-Doucas-Kimmitt}
G.~Doucas, M.~F. Kimmitt, A.~Doria, G.~P. Gallerano, E.~Giovenale, G.~Messina,
  H.~L. Andrews, and J.~H. Brownell, \enquote{Determination of longitudinal
  bunch shape by means of coherent {Smith-Purcell} radiation,}
  {\protect\JournalTitle{Physical Review Special Topics - Accelerators and
  Beams}} \textbf{5} (2002).

\bibitem{2009-Blackmore-Doucas}
V.~Blackmore, G.~Doucas, C.~Perry, B.~Ottewell, M.~F. Kimmitt, M.~Woods,
  S.~Molloy, and R.~Arnold, \enquote{First measurements of the longitudinal
  bunch profile of a 28.5~{GeV} beam using coherent {Smith-Purcell} radiation,}
  {\protect\JournalTitle{Physical Review Special Topics - Accelerators and
  Beams}} \textbf{12} (2009).

\bibitem{2012-Bartolini-Clarke}
R.~Bartolini, C.~Clarke, N.~Delerue, G.~Doucas, and A.~Reichold,
  \enquote{Electron bunch profile reconstruction in the few fs regime using
  coherent {Smith}-{Purcell} radiation,} {\protect\JournalTitle{Journal of
  Instrumentation}} \textbf{7}, P01009--P01009 (2012).

\bibitem{2014-England-Noble}
R.~J. England, R.~J. Noble, K.~Bane, D.~H. Dowell, C.-K. Ng, J.~E. Spencer,
  S.~Tantawi, Z.~Wu, R.~L. Byer, E.~Peralta, K.~Soong, C.-M. Chang,
  B.~Montazeri, S.~J. Wolf, B.~Cowan, J.~Dawson, W.~Gai, P.~Hommelhoff, Y.-C.
  Huang, C.~Jing, C.~McGuinness, R.~B. Palmer, B.~Naranjo, J.~Rosenzweig,
  G.~Travish, A.~Mizrahi, L.~Schachter, C.~Sears, G.~R. Werner, and R.~B.
  Yoder, \enquote{Dielectric laser accelerators,}
  {\protect\JournalTitle{Reviews of Modern Physics}} \textbf{86}, 1337--1389
  (2014).

\bibitem{2020-Sapra-Yang}
N.~V. Sapra, K.~Y. Yang, D.~Vercruysse, K.~J. Leedle, D.~S. Black, R.~J.
  England, L.~Su, R.~Trivedi, Y.~Miao, O.~Solgaard, R.~L. Byer, and
  J.~Vu{\v{c}}kovi{\'{c}}, \enquote{On-chip integrated laser-driven particle
  accelerator,} {\protect\JournalTitle{Science}} \textbf{367}, 79--83 (2020).

\bibitem{2018-Yang-Massuda}
Y.~Yang, A.~Massuda, C.~Roques-Carmes, S.~E. Kooi, T.~Christensen, S.~G.
  Johnson, J.~D. Joannopoulos, O.~D. Miller, I.~Kaminer, and
  M.~Solja{\v{c}}i{\'{c}}, \enquote{Maximal spontaneous photon emission and
  energy loss from free electrons,} {\protect\JournalTitle{Nature Physics}}
  \textbf{14}, 894--899 (2018).

\bibitem{2010-Sukhikh-Naumenko}
L.~G. Sukhikh, G.~A. Naumenko, Y.~A. Popov, and A.~P. Potylitsyn,
  \enquote{Comparison of coherent {S}mith-{P}urcell radiation generated by 6.1
  {MeV} electron beam in metal and dielectric lamellar gratings,}  (2010).

\bibitem{2011-Potylitsyn}
A.~P. Potylitsyn, \emph{{S}mith--{P}urcell Radiation} (Springer Berlin
  Heidelberg, Berlin, Heidelberg, 2011), pp. 135--164.

\bibitem{2006-Karlovets-Potylitsyn}
D.~V. Karlovets and A.~P. Potylitsyn, \enquote{Comparison of {S}mith-{P}urcell
  radiation models and criteria for their verification,}
  {\protect\JournalTitle{Physical Review Special Topics - Accelerators and
  Beams}} \textbf{9} (2006).

\bibitem{2016-Malovytsia-Delerue}
M.~Malovytsia and N.~Delerue, \enquote{{C}omparison of the {S}mith{-}{P}urcell
  {R}adiation {Y}ield for {D}ifferent {M}odels,} in \emph{Proc. of
  International Particle Accelerator Conference (IPAC'16), Busan, Korea, May
  8-13, 2016,}  (JACoW, Geneva, Switzerland, 2016), no.~7 in International
  Particle Accelerator Conference, pp. 75--77.
  Doi:10.18429/JACoW-IPAC2016-MOPMB004.

\bibitem{1960-diFrancia}
G.~T. di~Francia, \enquote{On the theory of some \v{C}erenkovian effects,}
  {\protect\JournalTitle{Il Nuovo Cimento}} \textbf{16}, 61--77 (1960).

\bibitem{1966-Barnes-Dedrick}
C.~W. Barnes and K.~G. Dedrick, \enquote{Radiation by an electron beam
  interacting with a diffraction grating. two-dimensional theory,}
  {\protect\JournalTitle{Journal of Applied Physics}} \textbf{37}, 411--418
  (1966).

\bibitem{1973-Lalor}
{\'{E}}.~Lalor, \enquote{Three-dimensional theory of the
  {S}mith{\textemdash}{P}urcell effect,} {\protect\JournalTitle{Physical Review
  A}} \textbf{7}, 435--446 (1973).

\bibitem{1998-Brownell-Walsh}
J.~H. Brownell, J.~Walsh, and G.~Doucas, \enquote{Spontaneous {S}mith-{P}urcell
  radiation described through induced surface currents,}
  {\protect\JournalTitle{Physical Review E}} \textbf{57}, 1075--1080 (1998).

\bibitem{2000-Trotz-Brownell}
S.~R. Trotz, J.~H. Brownell, J.~E. Walsh, and G.~Doucas, \enquote{Optimization
  of {S}mith-{P}urcell radiation at very high energies,}
  {\protect\JournalTitle{Physical Review E}} \textbf{61}, 7057--7064 (2000).

\bibitem{2002-Kube-Backe}
G.~Kube, H.~Backe, H.~Euteneuer, A.~Grendel, F.~Hagenbuck, H.~Hartmann, K.~H.
  Kaiser, W.~Lauth, H.~Sch\"{o}pe, G.~Wagner, T.~Walcher, and M.~Kretzschmar,
  \enquote{Observation of optical {S}mith-{P}urcell radiation at an electron
  beam energy of 855 {MeV},} {\protect\JournalTitle{Physical Review E}}
  \textbf{65} (2002).

\bibitem{2005-Brownell-Doucas}
J.~H. Brownell and G.~Doucas, \enquote{Role of the grating profile in
  {S}mith-{P}urcell radiation at high energies,}
  {\protect\JournalTitle{Physical Review Special Topics - Accelerators and
  Beams}} \textbf{8} (2005).

\bibitem{1973-vandenBerg2D}
P.~M. van~den Berg, \enquote{{S}mith{\textendash}{P}urcell radiation from a
  line charge moving parallel to a reflection grating,}
  {\protect\JournalTitle{Journal of the Optical Society of America}}
  \textbf{63}, 689 (1973).

\bibitem{1973-vandenBerg3D}
P.~M. van~den Berg, \enquote{{S}mith{\textendash}{P}urcell radiation from a
  point charge moving parallel to a reflection grating,}
  {\protect\JournalTitle{Journal of the Optical Society of America}}
  \textbf{63}, 1588 (1973).

\bibitem{1974-vandenBerg-Tan}
P.~M. van~den Berg and T.~H. Tan, \enquote{{S}mith-{P}urcell radiation from a
  line charge moving parallel to a reflection grating with rectangular
  profile,} {\protect\JournalTitle{Journal of the Optical Society of America}}
  \textbf{64}, 325 (1974).

\bibitem{1998-Shibata-Hasebe}
Y.~Shibata, S.~Hasebe, K.~Ishi, S.~Ono, M.~Ikezawa, T.~Nakazato, M.~Oyamada,
  S.~Urasawa, T.~Takahashi, T.~Matsuyama, K.~Kobayashi, and Y.~Fujita,
  \enquote{Coherent {S}mith-{P}urcell radiation in the millimeter-wave region
  from a short-bunch beam of relativistic electrons,}
  {\protect\JournalTitle{Physical Review E}} \textbf{57}, 1061--1074 (1998).

\bibitem{1994-Haeberle-Rullhusen}
O.~Haeberl{\'{e}}, P.~Rullhusen, J.-M. Salom{\'{e}}, and N.~Maene,
  \enquote{Calculations of {S}mith-{P}urcell radiation generated by electrons
  of 1{\textendash}100 {MeV},} {\protect\JournalTitle{Physical Review E}}
  \textbf{49}, 3340--3352 (1994).

\bibitem{2017-Lai-Kuang}
Y.-C. Lai, T.~C. Kuang, B.~H. Cheng, Y.-C. Lan, and D.~P. Tsai,
  \enquote{Generation of convergent light beams by using surface plasmon locked
  {Smith}-{Purcell} radiation,} {\protect\JournalTitle{Scientific Reports}}
  \textbf{7} (2017).

\bibitem{2017-Kaminer-Kooi}
I.~Kaminer, S.~Kooi, R.~Shiloh, B.~Zhen, Y.~Shen, J.~L{\'{o}}pez, R.~Remez,
  S.~Skirlo, Y.~Yang, J.~Joannopoulos, A.~Arie, and M.~Solja{\v{c}}i{\'{c}},
  \enquote{Spectrally and spatially resolved {Smith}-{Purcell} radiation in
  plasmonic crystals with short-range disorder,}
  {\protect\JournalTitle{Physical Review X}} \textbf{7} (2017).

\bibitem{2017-Remez-Shapira}
R.~Remez, N.~Shapira, C.~Roques-Carmes, R.~Tirole, Y.~Yang, Y.~Lereah,
  M.~Solja{\v{c}}i{\'{c}}, I.~Kaminer, and A.~Arie, \enquote{Spectral and
  spatial shaping of {Smith}-{Purcell} radiation,}
  {\protect\JournalTitle{Physical Review A}} \textbf{96} (2017).

\bibitem{2018-Massuda-Roques-Carmes}
A.~Massuda, C.~Roques-Carmes, Y.~Yang, S.~E. Kooi, Y.~Yang, C.~Murdia, K.~K.
  Berggren, I.~Kaminer, and M.~Solja{\v{c}}i{\'{c}},
  \enquote{{S}mith{\textendash}{P}urcell radiation from low-energy electrons,}
  {\protect\JournalTitle{{ACS} Photonics}} \textbf{5}, 3513--3518 (2018).

\bibitem{2014-Liu-Xu}
W.~Liu and Z.~Xu, \enquote{Special {S}mith{\textendash}{P}urcell radiation from
  an open resonator array,} {\protect\JournalTitle{New Journal of Physics}}
  \textbf{16}, 073006 (2014).

\bibitem{2015-Lekomtsev-Karataev}
K.~Lekomtsev, P.~Karataev, A.~Tishchenko, and J.~Urakawa, \enquote{{CST}
  simulations of {THz} {S}mith{\textendash}{P}urcell radiation from a lamellar
  grating with vacuum gaps,} {\protect\JournalTitle{Nuclear Instruments and
  Methods in Physics Research Section B: Beam Interactions with Materials and
  Atoms}} \textbf{355}, 164--169 (2015).

\bibitem{2017-Aryshev-Potylitsyn}
A.~Aryshev, A.~Potylitsyn, G.~Naumenko, M.~Shevelev, K.~Lekomtsev, L.~Sukhikh,
  P.~Karataev, Y.~Honda, N.~Terunuma, and J.~Urakawa, \enquote{Monochromaticity
  of coherent {Smith}-{Purcell} radiation from finite size grating,}
  {\protect\JournalTitle{Physical Review Accelerators and Beams}} \textbf{20}
  (2017).

\bibitem{2019-Zhang-Konoplev}
H.~Zhang, I.~Konoplev, and G.~Doucas, \enquote{A coherent {Smith}-{Purcell}
  radiation source: design considerations for a high power, tunable source of
  terahertz radiation,} in \emph{2019 44th International Conference on
  Infrared, Millimeter, and Terahertz Waves ({IRMMW}-{THz}),}  ({IEEE}, 2019).

\bibitem{2018-Song-Du}
Y.~Song, J.~Du, N.~Jiang, L.~Liu, and X.~Hu, \enquote{Efficient terahertz and
  infrared {S}mith{\textendash}{P}urcell radiation from metal-slot
  metasurfaces,} {\protect\JournalTitle{Optics Letters}} \textbf{43}, 3858
  (2018).

\bibitem{2018-Tyukhtin-Vorobev}
A.~Tyukhtin, V.~Vorobev, E.~Belonogaya, and S.~Galyamin, \enquote{Radiation of
  a charge in presence of a dielectric object: aperture method,}
  {\protect\JournalTitle{Journal of Instrumentation}} \textbf{13},
  C02033--C02033 (2018).

\bibitem{2018-Galyamin-Tyukhtin}
S.~Galyamin, A.~Tyukhtin, and V.~Vorobev, \enquote{Focusing the cherenkov
  radiation using dielectric concentrator: simulations and comparison with
  theory,} {\protect\JournalTitle{Journal of Instrumentation}} \textbf{13},
  C02029--C02029 (2018).

\bibitem{2019-Galyamin-Vorobev}
S.~N. Galyamin, V.~V. Vorobev, and A.~Benediktovitch, \enquote{Radiation field
  of an ideal thin gaussian bunch moving in a periodic conducting wire
  structure,} {\protect\JournalTitle{Physical Review Accelerators and Beams}}
  \textbf{22} (2019).

\bibitem{Comsol}
\enquote{{Comsol Multiphysics, version 4.3},} \url{https://www.comsol.com}.

\bibitem{1937-Frank-Tamm}
I.~Frank and I.~Tamm, \enquote{{Coherent visible radiation of fast electrons
  passing through matter},} {\protect\JournalTitle{Comptes rendus de
  l'Acad\'{e}mie des sciences de l'{URSS}}} \textbf{{14}}, {109--114} ({1937}).

\bibitem{1998-Jackson}
J.~D. Jackson, \emph{Classical Electrodynamics} (Wiley, 1998), 3rd ed.

\bibitem{2004-Shiozawa}
T.~Shiozawa, \emph{Classical Relativistic Electrodynamics} (Springer Berlin
  Heidelberg, 2004).

\bibitem{2015-Liu-Li}
W.~Liu, W.~Li, Z.~He, and Q.~Jia, \enquote{Theory of the special
  {S}mith-{P}urcell radiation from a rectangular grating,}
  {\protect\JournalTitle{{AIP} Advances}} \textbf{5}, 127135 (2015).

\bibitem{1984-Haus}
H.~Haus, \emph{Waves and fields in optoelectronics} (Prentice Hall,
  Incorporated, 1984).

\bibitem{2011-Schachter}
L.~Sch\"{a}chter, \emph{Beam-Wave Interaction in Periodic and Quasi-Periodic
  Structures} (Springer Berlin Heidelberg, 2011).

\bibitem{1984-Chuang-Kong}
S.~L. Chuang and J.~A. Kong, \enquote{Enhancement of
  {S}mith{\textendash}{P}urcell radiation from a grating with surface-plasmon
  excitation,} {\protect\JournalTitle{Journal of the Optical Society of America
  A}} \textbf{1}, 672 (1984).

\bibitem{2018-Pan-Gover}
Y.~Pan and A.~Gover, \enquote{Spontaneous and stimulated radiative emission of
  modulated free-electron quantum wavepackets{\textemdash}semiclassical
  analysis,} {\protect\JournalTitle{Journal of Physics Communications}}
  \textbf{2}, 115026 (2018).

\bibitem{2019-Szczepkowicz}
A.~Szczepkowicz, \enquote{Numerical calculation of the {S}mith-{P}urcell
  radiation from dielectric laser acceleration ({DLA}) structures,}
  \url{https://agenda.infn.it/event/17304/contributions/98899/} (2019). Poster
  presented at the European Advanced Accelerator Concepts Workshop (EAAC2019),
  Elba, Italy.

\bibitem{2016-Szczepkowicz}
A.~Szczepkowicz, \enquote{Guided-mode resonance, resonant grating thickness,
  and finite-size effects in dielectric laser acceleration structures,}
  {\protect\JournalTitle{Applied Optics}} \textbf{55}, 2634 (2016).

\bibitem{2020-Hausler}
U.~H{\"{a}}usler \emph{et~al.} In preparation.

\bibitem{2015-McPeak-Jayanti}
K.~M. McPeak, S.~V. Jayanti, S.~J.~P. Kress, S.~Meyer, S.~Iotti, A.~Rossinelli,
  and D.~J. Norris, \enquote{Plasmonic films can easily be better: Rules and
  recipes,} {\protect\JournalTitle{ACS Photonics}} \textbf{2}, 326--333 (2015).
  PMID: 25950012.

\bibitem{refractiveindexinfo}
M.~Polyanskiy, \enquote{Refractive index database,}
  \url{https://refractiveindex.info/} (2020).

\bibitem{1965-Malitson}
I.~H. Malitson, \enquote{Interspecimen comparison of the refractive index of
  fused silica,} {\protect\JournalTitle{Journal of the Optical Society of
  America}} \textbf{55}, 1205 (1965).

\bibitem{2015-Schinke-Peest}
C.~Schinke, P.~C. Peest, J.~Schmidt, R.~Brendel, K.~Bothe, M.~R. Vogt,
  I.~Kr\"{o}ger, S.~Winter, A.~Schirmacher, S.~Lim, H.~T. Nguyen, and
  D.~MacDonald, \enquote{Uncertainty analysis for the coefficient of
  band-to-band absorption of crystalline silicon,} {\protect\JournalTitle{{AIP}
  Advances}} \textbf{5}, 067168 (2015).

\bibitem{2019-Yousefi-Schonenberger}
P.~Yousefi, N.~Sch\"{o}nenberger, J.~Mcneur, M.~Koz{\'{a}}k, U.~Niedermayer,
  and P.~Hommelhoff, \enquote{Dielectric laser electron acceleration in a dual
  pillar grating with a distributed bragg reflector,}
  {\protect\JournalTitle{Optics Letters}} \textbf{44}, 1520 (2019).

\bibitem{2011-Potylitsyn-Ryazanov}
A.~P. Potylitsyn, M.~I. Ryazanov, M.~N. Strikhanov, and A.~A. Tishchenko,
  \emph{Diffraction Radiation from Relativistic Particles} (Springer Berlin
  Heidelberg, 2011).

\bibitem{1981-Nosich}
A.~I. Nosich, \enquote{Diffraction radiation which accompanies the motion of
  charged particles near an open resonator,}
  {\protect\JournalTitle{Radiophysics and Quantum Electronics}} \textbf{24},
  696--701 (1981).

\bibitem{2019-Yevtushenko-Dukhopelnykov}
D.~O. Yevtushenko, S.~V. Dukhopelnykov, and A.~I. Nosich, \enquote{Optical
  diffraction radiation from a dielectric and a metal nanowire excited by a
  modulated electron beam,} {\protect\JournalTitle{Optical and Quantum
  Electronics}} \textbf{51} (2019).

\bibitem{1977-Veliev-Nosich}
{\'{E}}.~I. Veliev, A.~I. Nosich, and V.~P. Shestopalov, \enquote{Radiation of
  an electron flux moving over a grating consisting of cylinders with
  longitudinal slits,} {\protect\JournalTitle{Radiophysics and Quantum
  Electronics}} \textbf{20}, 306--313 (1977).

\end{thebibliography}


\end{document}